\documentclass[conference]{IEEEtran}
\usepackage{graphicx} 
\usepackage{amsmath,amssymb,amsfonts,amsthm}
\usepackage{multicol,lipsum}
\usepackage{mathtools}
\usepackage{cite}
\usepackage{algorithmic}
\usepackage{algorithm}
\usepackage{enumerate}   

\DeclareMathOperator{\Tr}{Tr}
\DeclareMathOperator{\Vect}{vec}

\usepackage{subcaption}
\usepackage{graphicx}
\usepackage{textcomp}
\usepackage{optidef}
\usepackage{xcolor}

\title{Integrated Sensing, Computing, and Semantic Communication with Fluid Antenna for Metaverse}

\begin{document}
\author{Yinchao Yang, Jingxuan Zhou,
Zhaohui Yang
\thanks{Yinchao Yang and Jingxuan Zhouare with the Department of Engineering, King's College London, London, UK. (emails: yinchao.yang@kcl.ac.uk; jingxuan.zhou@kcl.ac.uk)}
\thanks{Zhaohui Yang is with the College of Information Science and Electronic Engineering, Zhejiang University, Hangzhou, Zhejiang 310027, China, and Zhejiang Provincial Key Lab of Information Processing, Communication and Networking (IPCAN), Hangzhou, Zhejiang, 310007, China. (email: yang\_zhaohui@zju.edu.cn) }}

\author{
    \IEEEauthorblockN{Yinchao Yang \IEEEauthorrefmark{1}, 
    Jingxuan Zhou \IEEEauthorrefmark{1},
    Zhaohui Yang \IEEEauthorrefmark{2}\IEEEauthorrefmark{3}}
    \IEEEauthorblockA{
    $\IEEEauthorrefmark{1}$Department of Engineering, King's College London, London, UK \\
    $\IEEEauthorrefmark{2}$College of Information Science and 
    Electronic Engineering, Zhejiang University, Hangzhou, China\\
    $\IEEEauthorrefmark{3}$Zhejiang Provincial Key Laboratory of Info. Proc., Commun. \& Netw. (IPCAN), Hangzhou, China\\
    E-mails: 
    yinchao.yang@kcl.ac.uk,
    jingxuan.zhou@kcl.ac.uk,
    yang\_zhaohui@zju.edu.cn
    }
}

\maketitle

\IEEEpeerreviewmaketitle
\begin{abstract}


The integration of sensing and communication (ISAC) is pivotal for the Metaverse but faces challenges like high data volume and privacy concerns. This paper proposes a novel integrated sensing, computing, and semantic communication (ISCSC) framework, which uses semantic communication to transmit only contextual information, reducing data overhead and enhancing efficiency. To address the sensitivity of semantic communication to channel conditions, fluid antennas (FAs) are introduced, enabling dynamic adaptability. The FA-enabled ISCSC framework considers multiple users and extended targets composed of a series of scatterers, formulating a joint optimization problem to maximize the data rate while ensuring sensing accuracy and meeting computational and power constraints. An alternating optimization (AO) method decomposes the problem into subproblems for ISAC beamforming, FA positioning, and semantic extraction. Simulations confirm the framework’s effectiveness in improving data rates and sensing performance.
\end{abstract}

\begin{IEEEkeywords}
Integrated sensing and communication, semantic communication, transmit beamforming, fluid antennas, Metaverse.
\end{IEEEkeywords}

\section{Introduction}


As emerging applications like virtual and augmented reality (VR/AR) and the Metaverse continue to expand, the integration of wireless sensing and communication is set to play a crucial role in enabling next-generation services. The Metaverse, a virtual shared space that seamlessly blends physical and digital realities, depends on cutting-edge technologies to provide immersive experiences, making the integration of sensing and communication essential \cite{wang2023semantic, lotfi2023semantic, ding2024joint, 10251844, bobarshad2009m,4133008,fang2021secure,nehra2010cross,shadmand2010multi, jia2020channel}. Recognizing the transformative potential of this approach, the International Telecommunication Union (ITU) has identified integrated sensing and communication (ISAC) as one of the six core usage scenarios in its global vision for sixth-generation (6G) mobile communication systems. ISAC operates on the principle of utilizing shared wireless resources—such as power, frequency bands, beams, and hardware infrastructure—to deliver both sensing and communication capabilities \cite{liu2022integrated}. Research has shown that ISAC systems can surpass standalone sensing and communication solutions in terms of resource efficiency and can achieve synergistic benefits, where sensing enhances communication and communication supports sensing.

Although ISAC offers significant advantages, it also faces notable challenges, particularly in handling the vast volumes of data transmitted and received in the Metaverse. This data deluge can result in increased latency and computational overhead, ultimately degrading system performance \cite{gu2023semantic,yang2023secure}. Another critical issue pertains to data privacy risks, especially in the context of regulations such as the General Data Protection Regulation (GDPR). When sensing and communication signals are transmitted simultaneously, unintended users (e.g., sensing targets) may inadvertently receive sensitive communication messages, raising serious privacy concerns \cite{zhaohui2024secure}. Integrating semantic communication into the ISAC framework offers a promising solution to these challenges. Unlike conventional communication methods, semantic communication transmits the meaning and relevance of information rather than raw data, enabling intelligent prioritization and compression based on contextual importance \cite{luo2022semantic}. This requires both transmitters and receivers to establish shared knowledge bases (KBs) containing essential information accessible to both parties. This approach not only reduces data overhead but also enhances privacy by ensuring that unintended users without appropriate KBs cannot decode the transmitted semantic content. Furthermore, semantic communication supports efficient multi-modal data transmission, meeting the diverse requirements of various communication users \cite{zhang2024unified}, particularly in the Metaverse.

Integrating semantic communication into the ISAC framework, however, is not without its challenges. Semantic communication is inherently more sensitive to channel conditions than traditional systems. While conventional communication focuses on ensuring bit-level accuracy, semantic communication prioritizes preserving the meaning or context of transmitted data \cite{wang2025generative, xu2023edge, yang2023energy}. As a result, poor channel conditions can distort the semantic content of a message, potentially leading to misunderstandings or incorrect decisions by the receiver. A promising solution to stabilize channel conditions is the use of fluid antennas (FAs), which can dynamically adjust their positions to reconfigure radiation characteristics \cite{ghadi2024physical, zhou2024near}. Unlike fixed-position antennas (FPAs), FAs offer a higher degree of freedom (DoF) to adapt to varying channel conditions by effectively exploring channel variations. Research has demonstrated the significant advantages of FAs in wireless communication systems. For example, Zhu \textit{et al.} \cite{zhu2023modeling} showed that FAs outperform FPAs, particularly in environments with an increasing number of channel paths, as they can better leverage small-scale fading effects in the spatial domain.

In our previous work \cite{yang2024secure,yang2024joint}, we introduced the framework of integrated sensing, computing, and semantic communication (ISCSC). However, the integration of fluid antennas (FAs) into this system and their potential benefits were not explored, leaving a notable research gap. This gap serves as the foundation for developing \textbf{FA-enabled ISCSC} systems. Building on this motivation, the key contributions of this paper are as follows:
\begin{enumerate}
    \item \textbf{Modeling FA-Enabled ISCSC for the Metaverse}:
   We propose a novel FA-enabled ISCSC framework tailored for the Metaverse. Unlike conventional ISAC, the integration of semantic communication not only enhances communication efficiency but also improves data privacy. Additionally, the deployment of FAs allows dynamic control over channel conditions, thereby stabilizing and optimizing semantic communication performance.

    \item \textbf{Accurate Target Modelling}:  
    This work adopts an extended target model, where each target is represented by multiple scatterers, offering a detailed characterization of real-world physical properties. This precise modelling is particularly critical for the Metaverse, where accurate and high-fidelity object representations are essential for immersive simulations and informed decision-making.

\end{enumerate}

\section{System Model}

We investigate an FA-enabled ISCSC system with a single BS, where the transmit and receive antennas are co-located to facilitate both target detection and downlink semantic communication. In this setup, the BS is equipped with a planar array of $N_{\text{tx}} \times N_{\text{tz}}$ FAs for signal transmission, while $N_{\text{rx}} \times N_{\text{rz}}$ FPAs are used for signal reception. Each FA, indexed by $i \in \{1, \dots, N_{\text{tx}} \times N_{\text{tz}}\}$, is associated with a spatial coordinate $\mathbf{u}_i = [x_i, z_i]$ and is capable of dynamic movement within a defined rectangular region $\mathcal{C}_i = [x_i^{\text{min}}, x_i^{\text{max}}] \times [z_i^{\text{min}}, z_i^{\text{max}}]$. In contrast, each FPA, indexed by $m \in \{1, \dots, N_{\text{rx}} \times N_{\text{rz}}\}$, is fixed at a specific coordinate $\mathbf{v}_m = [x_m, z_m]$. 

The BS communicates with $K$ communication users (CUs), and each CU $k \in \mathcal{K}$ is equipped with a single antenna. Simultaneously, the BS actively detects $L$ targets, each target $l \in \mathcal{L}$ being considered as an extended target.

\begin{figure}[!t]
    \centering
    \includegraphics[width=0.9\linewidth]{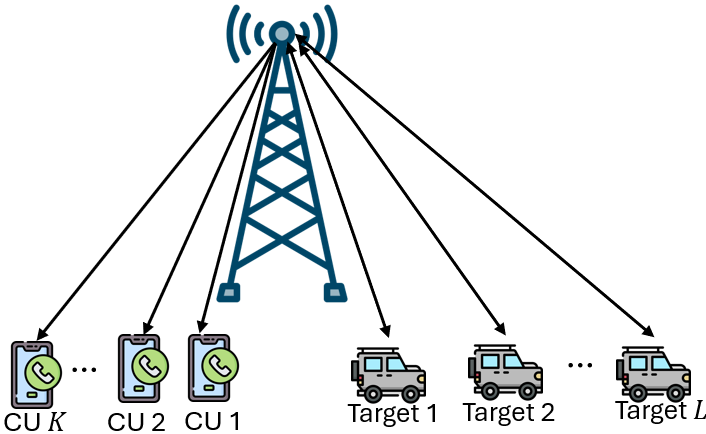}
    \caption{A FA-enabled ISCSC system with multiple users and multiple targets.}
    \label{FA geometry}
\end{figure}

\subsection{Signal Model}

In the proposed framework, the BS simultaneously transmits semantic signals to the CUs and sensing signals to the targets through the utilization of beamforming. The transmitted signal $\mathbf{X} \in \mathbb{C}^{(N_{\text{tx}} \times N_{\text{tz}}) \times F}$, which represents the joint signal with $F > N_{\text{tx}} \times N_{\text{tz}}$ frames, can be expressed as
\begin{equation}\label{ISCSC signal} 
\mathbf{X} = \mathbf{W} \mathbf{C} + \mathbf{R}, 
\end{equation} 
where $\mathbf{W} = [\mathbf{w}_1, \dots, \mathbf{w}_K]$ denotes the communication beamforming matrix for the CUs, with $\mathbf{w}_k \in \mathbb{C}^{(N_{\text{tx}} \times N_{\text{tz}}) \times 1}$ representing the beamforming vector for each user. The communication message intended for the CUs is denoted by $\mathbf{C} = [\mathbf{c}_1, \dots, \mathbf{c}_K]^H$, where $\mathbf{c}_k \in \mathbb{C}^{1 \times F}$ represents the communication signal for each CU. Additionally, $\mathbf{R} \in \mathbb{C}^{(N_{\text{tx}} \times N_{\text{tz}}) \times F}$ represents the sensing beamforming matrix. The covariance matrix of the transmitted signal $\mathbf{X}$ is given by
\begin{equation} \label{covar}
    \mathbf{R}_x = \mathbb{E} \left[\mathbf{X}\mathbf{X}^H \right] = \mathbf{W} \mathbf{W}^H + \mathbf{R} \mathbf{R}^H.
\end{equation}

\subsection{Communication Model}
After the BS transmits the signal $\mathbf{X}$, the received signal at the $k$-th CU can be expressed as: 
\begin{equation}\label{user received} 
    \mathbf{y}_k = \alpha_k \mathbf{a}_t^H\left(\theta_k, \phi_k, \mathbf{u} \right) \mathbf{X} + \mathbf{n}_k = \mathbf{h}_k^H \mathbf{X} + \mathbf{n}_k, 
\end{equation} 
where $\alpha_k$ represents the path-loss coefficient for the $k$-th user, and $\mathbf{a}_t\left(\theta_k, \phi_k,\mathbf{u} \right) \in\mathbb{C}^{(N_{\text{tx}} \times N_{\text{tz}}) \times 1}$ denotes the steering vector, $\theta_k$ and $\phi_k$ represent the azimuth angle and the broadside angle, respectively. Additionally, $\mathbf{n}_k \sim \mathcal{CN}(0,\sigma^2_c \mathbf{I}_{1 \times F})$ represents the communication noise.

On the target side, the received signal can be modelled as: 
\begin{equation}\label{eve receive} 
    \mathbf{y}_l = \sum_{s=1}^{N_s} \alpha_{l,s} \mathbf{a}_t^H \left(\theta_{l,s}, \phi_{l,s}, \mathbf{u} \right) \mathbf{X} + \mathbf{n}_l = \mathbf{h}_l^H \mathbf{X} + \mathbf{n}_l, 
\end{equation} 
where $N_s$ is the number of scatterers forming an extended target, $\alpha_{l}$ is the path-loss coefficient, and $\mathbf{a}_t\left(\theta_{l}, \phi_{l}, \mathbf{u} \right) \in \mathbb{C}^{(N_{\text{tx}} \times N_{\text{tz}}) \times 1}$ denotes the corresponding steering vector for the target. The term $\mathbf{n}_l \sim \mathcal{CN} \left(0, \sigma^2_c \mathbf{I}_{1 \times F} \right)$ represents the noise encountered by the target. The steering vector can be characterized by \cite{liu2023near}:
\begin{equation}\label{eq1}
        \mathbf{a}_t \left(\theta_q, \phi_q, \mathbf{u}\right) = \mathbf{a}_{\text{tx}}\left(\theta_q, \phi_q, \mathbf{x}_t \right) \otimes \mathbf{a}_{\text{tz}}\left(\phi_q, \mathbf{z}_t \right), q \in [k,l],
\end{equation}
where $\mathbf{a}_{\text{tx}}(\theta_q, \phi_q, \mathbf{x}_t) \in \mathbb{C}^{N_{\text{tx}} \times 1}$ and $\mathbf{a}_{\text{tz}}(\phi_q, \mathbf{z}_t) \in \mathbb{C}^{N_{\text{tz}} \times 1}$. The $x$-coordinates and $z$-coordinates of all FAs are represented by $\mathbf{x}_t$ and $\mathbf{z}_t$, respectively. Each element in the steering vector can be calculated using the following equations:
\begin{equation}\label{eq1.2}
\begin{aligned}
    & a_{\text{tx}}\left(\theta_q, \phi_q, x_i \right) = e^{j\frac{2\pi}{\lambda} \left(x_i \cos\left(\theta_q\right) \sin\left(\phi_q\right)\right)}, \quad x_i \in \mathbf{x}_t, \\
    & a_{\text{tz}}\left(\phi_q, z_i \right) = e^{j\frac{2\pi}{\lambda} \left(z_i \cos\left(\phi_q \right) \right)}, \quad z_i \in \mathbf{z}_t.
\end{aligned}
\end{equation}

\subsection{Sensing Model}
After the BS transmits a joint signal to the locations of interest, the targets will reflect echo signals. The echo signal received by the BS, which contains information from all the targets, can be expressed as:
\begin{equation}\label{echo}
\begin{aligned}
    \mathbf{Z} &= \sum_{l \in \mathcal{L}} \sum_{s=1}^{N_s} \beta_{l,s} \mathbf{a}_r\left(\theta_{l,s}, \phi_{l,s}, \mathbf{v} \right) \mathbf{a}_t^H \left(\theta_{l,s}, \phi_{l,s}, \mathbf{u} \right) \mathbf{X} + \mathbf{N} \\
    &= \mathbf{G} \mathbf{X} + \mathbf{N},
\end{aligned}
\end{equation}
where $\mathbf{N} \sim \mathcal{CN}\left(0, \sigma^2_{r} \mathbf{I}_{(N_{\text{rx}} \times N_{\text{rz}}) \times F}\right)$ represents the Gaussian noise, and $\mathbf{a}_r(\theta_{l}, \phi_{l}, \mathbf{v}) \in \mathbb{C}^{(N_{\text{rx}} \times N_{\text{rz}}) \times 1}$ is the receiver steering vector, whose formulation is given below
\begin{equation}\label{eq2}
\begin{aligned}
    \mathbf{a}_r\left(\theta_{l}, \phi_{l}, \mathbf{v} \right) =\mathbf{a}_{\text{rx}}\left(\theta_{l}, \phi_{l}, \mathbf{x}_r \right) \otimes \mathbf{a}_{\text{rz}}\left(\phi_{l}, \mathbf{z}_r \right),
\end{aligned}
\end{equation}
where
\begin{equation}\label{eq2.2}
\begin{aligned}
    & a_{\text{rx}}\left(\theta_{l}, \phi_{l}, x_m \right) = e^{j\frac{2\pi}{\lambda} \left(m d_x \cos\left(\theta_{l} \right) \sin\left(\phi_{l} \right)\right)},    x_m \in \mathbf{x}_r,\\
    & a_{\text{rz}}\left(\phi_{l}, z_m \right) = e^{j\frac{2\pi}{\lambda} \left(m d_z \cos\left(\phi_{l} \right) \right)}, z_m \in \mathbf{z}_r,
\end{aligned}
\end{equation}
where $\mathbf{x}_r$ and $\mathbf{z}_r$ contain the $x$-coordinates and $z$-coordinates of all the FPAs, respectively. The x-axis and z-axis antenna spacing are represented by $d_x$ and $d_z$, respectively.

\section{Performance Measures}

\subsection{Sensing} 
The Cramér-Rao bound (CRB) serves as a benchmark for the minimum achievable mean square error, defining the best possible estimation accuracy for a system under ideal conditions. For extended targets, we estimate the entire echo channel matrix $\mathbf{G}$. Accurate estimation of the channel matrix enables the use of algorithms for extracting relevant information.

To compute the CRB, we first need to calculate the Fisher information matrix (FIM), which quantifies the amount of information contained in an observed variable about the unknown parameters of interest. By vectorizing $\mathbf{Z}$ and denoting it as $\mathbf{\bar{z}}$, we obtain:
\begin{equation}
    \mathbf{\bar{z}}= \Vect \left({\mathbf{Z}} \right) = \left( \mathbf{X}^T \otimes \mathbf{I}_{\left(N_{\text{tx}} \times N_{\text{tz}}\right)} \right) \mathbf{\bar{g}} + \mathbf{\bar{n}},
\end{equation}
where $\mathbf{\bar{g}} = \Vect \left( \mathbf{G}\right)$ and $\mathbf{\bar{n}} = \Vect \left( \mathbf{N}\right)$. According to \cite{liu2021cramer}, the FIM of $\mathbf{\bar{g}}$ is give by
\begin{equation}\label{fim}
    \mathbf{J} = \frac{1}{\sigma^2_r} \mathbf{X}^* \mathbf{X}^T \otimes \mathbf{I}_{\left( N_{\text{tx}} \times N_{\text{tz}} \right) } = \frac{F}{\sigma_r^2} \mathbf{R}_x^T \otimes \mathbf{I}_{\left( N_{\text{tx}} \times N_{\text{tz}} \right)}.
\end{equation}

In \eqref{fim}, the rank of the matrix \(\mathbf{X}\) is \(N_{\text{tx}} \times N_{\text{tz}}\), which is sufficient to fully recover the channel matrix \(\mathbf{G}\), whose rank is also \(N_{\text{tx}} \times N_{\text{tz}}\). As emphasized in \cite{ben2009constrained, liu2021cramer}, a sufficient rank ensures that the FIM remains non-singular. Hence, the CRB of $\mathbf{\bar{g}}$ is given by 
\begin{equation}
    \text{CRB}\left(\mathbf{G} \right) = \mathbf{J}^{-1} = \frac{\sigma_r^2 N_{\text{rx}} N_{\text{rz}}}{F} \Tr\left( \mathbf{R}_x^{-1}\right).
\end{equation}

\subsection{Semantic Communication}
According to \eqref{user received}, the signal-to-noise-plus-interference (SINR) ratio of the $k$-th CU is given by
\begin{equation}\label{cu sinr}
    \gamma_k = \frac{\left| \mathbf{h}_k^H \mathbf{w}_k \right|^2}{\left|\mathbf{h}_k^H \sum_{k' \in \mathcal{K}, k' \neq k} \mathbf{w}_{k'} \right|^2 + \left| \left|\mathbf{h}_k^H \mathbf{R} \right| \right|^2 + \sigma_c^2}.
\end{equation}

The semantic transmission rate is defined as the number of bits received by the user after decoding the semantic information from the received signal. The expression for the semantic transmission rate is given by \cite{yang2024secure}:
\begin{equation}\label{semantic rate}
    R_k = \frac{\iota}{\rho_k} \log_2 \left( 1 + \gamma_k \right),
\end{equation}
where the parameter $\rho_{LB} \leq \rho_{k} \leq 1$ represents the semantic extraction ratio, and $\iota$ is a scalar value that converts the word-to-bit ratio. Additionally, $\rho_{LB}$ is the lower bound of $\rho_k$, with the formula provided in \cite[Lemma 1]{yang2024secure}.

In the joint transmission of sensing and communication signals, it is crucial to ensure that the unintended targets receive only a minimal amount of communication signal to prevent potential breaches of GDPR regulations. To quantify the extent of information intercepted by an unintended receiver, it is necessary to model the amount of information captured by target $l$ from the communication intended for user $k$. The semantic transmission rate of target $l$ is defined as:
\begin{equation}\label{eve rate}
    R_{l|k} = \frac{\iota}{\rho_k} \log_2 \left( 1 + \Gamma_{l|k} \right),
\end{equation}
where $\Gamma_{l|k}$ is the SINR of target $l$ related to CU $k$. And $\Gamma_{l|k}$ can be derived from \eqref{eve receive}, that is:
\begin{equation}\label{eve snr}
    \Gamma_{l | k} = \frac{\left| \mathbf{h}_l^H \mathbf{w}_k \right|^2}{ \left|\mathbf{h}_l^H \sum_{k' \in \mathcal{K}, k' \neq k} \mathbf{w}_{k'} \right|^2 + \left| \left|\mathbf{h}_l^H \mathbf{R} \right| \right|^2 + \sigma_c^2}.
\end{equation}

The semantic secrecy rate for the $k$-th CU can be formulated by incorporating \eqref{semantic rate} and \eqref{eve rate}:
\begin{equation}
    S_k = \min_{l \in \mathcal{L}} \left[ R_{k} - R_{l|k} \right]^{+}.
\end{equation}

\subsection{Computing}
Extracting semantic information from traditional messages requires computational resources. Therefore, it is crucial to consider computational power as a key component of the overall transmission power budget. As outlined in \cite{yang2024secure}, the computational power consumption is formulated by
\begin{equation}\label{eq18}
    P^{\text{Comp}} =  -\nu \sum_{k \in \mathcal{K}} \ln\left(\rho_{k}\right),
\end{equation}
where $\nu$ is a coefficient that converts the magnitude to its corresponding power. On the other hand, the power consumption for both communication and sensing at the BS is given by:
\begin{equation}\label{eq19}
    P^{\text{C\&S}} = \Tr\left(\mathbf{R}_x\right).
\end{equation}

\section{Joint Design of FA-enabled ISCSC System}

\subsection{Problem Formulation}
The design objective is to maximise the worst-case semantic secrecy rate, which simultaneously achieves the goal of enhancing the data rate. The optimisation problem is formulated as follows:
\begin{subequations} \label{opt1}
\begin{align}
    \max_{\mathbf{w}_k, \mathbf{R}_x, \mathbf{u}_i, \rho_k} \quad &   \min_{k \in \mathcal{K}} \left( S_k \right) \label{opt1a}\\
    \text{s.t.} \hspace{1cm} & \text{CRB}\left( \mathbf{G} \right) \leq \xi,  \label{opt1b}\\
    & P^{\text{C\&S}} + P^{\text{Comp}} \leq P_t,  \label{opt1c}\\
    & p_{LB} \leq \rho_{k} \leq 1,  \forall k, \label{opt1d} \\
    & \mathbf{u}_i \in C_i, \forall i, \label{opt1e}\\
    & \mathbf{R}_x \succeq \sum_{k \in \mathcal{K}} \mathbf{w}_k \mathbf{w}_k^H, \; \mathbf{w}_k \mathbf{w}_k^H \succeq 0,\forall k, \label{opt1f} \\
    & \text{rank} \left( \mathbf{w}_k \mathbf{w}_k^H\right) = 1, \forall k. \label{opt1g}
\end{align}
\end{subequations}

The constraint in \eqref{opt1b} guarantees the sensing performance by limiting the maximum CRB value to a predefined threshold. The constraint in \eqref{opt1c} bounds the total power consumption, encompassing the power allocated for signal transmission and semantic extraction, to stay within the maximum available transmission power. The semantic extraction ratio for each user, denoted as $\rho_k$, is constrained by \eqref{opt1d}, ensuring it remains between a lower bound, $p_{LB}$, and 1. Finally, the movement constraint in \eqref{opt1e} restricts the FA positions.

In the following section, we relax the rank-one constraint due to its non-convex nature. The rank-one solution can be recovered through Gaussian randomization. To overcome the non-convexity of the objective function in \eqref{opt1}, we propose an AO approach.

\subsection{Joint Beamforming and Computation Optimisation}
With given initial FA positions $\mathbf{u}_i$ and the initial semantic extraction ratios $\rho_k$, we can reformulate the original problem with respect to variables $[\mathbf{w}_k, \mathbf{R}_x, \zeta]$ as follows:
\begin{subequations}\label{opt2}
\begin{align}
    \max_{\mathbf{w}_k, \mathbf{R}_x, \zeta} \quad &   \zeta \label{opt2a}\\
    \text{s.t.}  \hspace{1cm} &  S_k \geq \zeta, \forall k, \label{opt2b} \\
    & \eqref{opt1b}, \eqref{opt1c}, \eqref{opt1f}.
\end{align}
\end{subequations}

In \eqref{opt2}, the first constraint is non-concave and presents a challenge for optimisation. To address this issue, we first reformulate \eqref{opt2b} as:
\begin{equation}\label{log expand}
    \log_2\left(A_k\right) - \log_2\left(B_k\right) + \log_2\left(C_{l|k}\right) - \log_2\left(D_{l|k}\right) \geq \zeta,
\end{equation}
where
\begin{equation}
    \begin{cases}
    \vspace{0.5cm}
    A_{k} = \left|\mathbf{h}_k^H \sum_{k \in \mathcal{K}} \mathbf{w}_{k} \right|^2 + \left| \left|\mathbf{h}_k^H \left(\mathbf{R}_x - \mathbf{w}_k \mathbf{w}_k^H \right) \right| \right|^2 + \sigma_c^2,\\

    B_{k} = \left|\mathbf{h}_k^H \sum_{k' \in \mathcal{K}, k' \neq k} \mathbf{w}_{k'} \right|^2 \\
    \vspace{0.5cm}
    \hspace{3cm} +\left| \left|\mathbf{h}_k^H \left(\mathbf{R}_x - \mathbf{w}_k \mathbf{w}_k^H \right) \right| \right|^2 + \sigma_c^2,\\

    C_{l|k} = \left|\mathbf{h}_l^H \sum_{k' \in \mathcal{K}, k' \neq k} \mathbf{w}_{k'} \right|^2 \\
    \vspace{0.5cm}
    \hspace{3cm} + \left| \left|\mathbf{h}_l^H\left(\mathbf{R}_x - \mathbf{w}_k \mathbf{w}_k^H \right) \right| \right|^2 + \sigma_c^2,\\
    \vspace{0.2cm}
    D_{l|k} = \left|\mathbf{h}_l^H \sum_{k \in \mathcal{K}} \mathbf{w}_{k} \right|^2 + \left| \left|\mathbf{h}_l^H\left(\mathbf{R}_x - \mathbf{w}_k \mathbf{w}_k^H \right) \right| \right|^2 + \sigma_c^2.
    \end{cases}
\end{equation}

The second and fourth terms in \eqref{log expand} remain non-convex. To handle this, we approximate these terms using a first-order Taylor expansion, resulting in the following expression:
\begin{equation}\label{log approxi}
\begin{cases}
    \vspace{0.5cm}
    \log_2\left(B_k\right) = \log_2\left(B_{k,e}\right) + \frac{1}{B_{k, e} \ln(2)} \left(B_k - B_{k, e}\right),\\
    \log_2\left(D_{l|k}\right) = \log_2\left(D_{l|k,e} \right) + \frac{1}{D_{l|k, e} \ln(2)} \left(D_{l|k} - D_{l|k, e}\right),\\
\end{cases}
\end{equation}
where the subscript $e$ denotes the value of the corresponding variable at each epoch. By substituting \eqref{log expand} and \eqref{log approxi}, the optimisation problem in \eqref{opt2} becomes convex and can be efficiently solved using established optimisation tools such as CVX \cite{grant2014cvx}.

\subsection{FA Position Optimisation}

For any given values of $\mathbf{w}_k$, $\mathbf{R}_x$, and $\rho_k$, the positions of the FAs $\mathbf{u}_i$ can be determined by solving the following optimisation problem:
\begin{subequations} \label{opt FA}
\begin{align}
    \max_{\mathbf{u}_i} \quad &  \min_{k \in \mathcal{K}} \left( S_k \right) \label{opt FA a}\\
    \text{s.t.} \quad &  \mathbf{u}_i \in C_i, \forall i. \label{opt FA b}
\end{align}
\end{subequations}

The optimisation problem in \eqref{opt FA} is non-convex due to the non-convex nature of the objective function in \eqref{opt FA a}. To address this, we use the second-order Taylor expansion to find a surrogate function. We first replace $\min_{k \in \mathcal{K}} \left( S_k \right)$ with the constraint $S_k \geq \xi, \forall k$, where $\xi$ is the auxiliary variable that we seek to maximise. To address the non-convexity of $S_k$, a second-order Taylor expansion is employed to construct a lower bound for $S_k$, which is shown in \eqref{log expansion obj}, where the Jacobian and Hessian matrices are denoted by $\nabla$ and $\nabla^2$, respectively. Additionally, the vector $\mathbf{u}$ is defined as $\mathbf{u} = [\mathbf{u}_1, \ldots, \mathbf{u}_i, \ldots, \mathbf{u}_{N_{\text{tx}} \times N_{\text{tz}}}]$.

\begin{table*}
\footnotesize
\centering
\begin{minipage}{1\textwidth}
    \begin{align} 
        &  g\left( \mathbf{u} \right) \overset{\Delta}{=}  \log_2\left(A_{k, e}\right) + \nabla\log_2\left(A_{k, e} \right) \left( \mathbf{u} - \mathbf{u}_e \right) - \frac{\nabla^2 \log_2 \left(A_{k, e}\right)}{2} \left( \mathbf{u} - \mathbf{u}_e \right)^T \left( \mathbf{u} - \mathbf{u}_e \right) -\log_2\left(B_{k, e}\right) - \nabla\log_2\left(B_{k, e}\right) \left( \mathbf{u} - \mathbf{u}_e \right) \nonumber \\
        &- \frac{\nabla^2 \log_2 \left(B_{k, e}\right)}{2} \left( \mathbf{u} - \mathbf{u}_e \right)^T \left( \mathbf{u} - \mathbf{u}^e \right) + \log_2\left(C_{l|k, e}\right) + \nabla\log_2\left(C_{l|k, e}\right) \left( \mathbf{u} - \mathbf{u}_e \right) - \frac{\nabla^2 \log_2 \left(C_{l|k, e}\right)}{2} \left( \mathbf{u} - \mathbf{u}_e \right)^T \left( \mathbf{u} - \mathbf{u}_e \right) \nonumber\\
        & -\log_2\left(D_{l|k, e}\right) - \nabla\log_2\left(D_{l|k, e}\right) \left( \mathbf{u} - \mathbf{u}_e \right) - \frac{\nabla^2 \log_2 \left(D_{l|k, e}\right)}{2} \left( \mathbf{u} - \mathbf{u}_e \right)^T \left( \mathbf{u} - \mathbf{u}_e \right). \label{log expansion obj}
    \end{align}
\medskip
\hrule
\end{minipage}
\end{table*}

By substituting each Hessian matrix in \eqref{log expansion obj} with suitable scalar values \([\delta_k, \epsilon_k, \delta_{l|k}, \epsilon_{l|k}]\), the optimisation problem in \eqref{opt FA} becomes convex and can be efficiently solved using existing optimisation tools. The computational complexity at each epoch is \(\mathcal{O} \left( \left( 2 N_{\text{tx}} N_{\text{tz}} \right)^3 \right)\), where \(2 N_{\text{tx}} N_{\text{tz}}\) represents the number of rows (or columns) of the Hessian matrix.

\subsection{Semantic Extraction Ratio Optimisation}
For any given values of $\mathbf{w}_k$, $\mathbf{R}_x$, and $\mathbf{u}_i$, the optimal values of $\rho_k$ can be determined by solving the following optimisation problem:
\begin{subequations} \label{opt semantic}
\begin{align}
    \min_{\rho_k} \quad &   \sum_{k \in \mathcal{K}} \rho_k \label{opt semantic a}\\
    \text{s.t.} \quad & P^{\text{Comp}} \leq P_t - P^{\text{C\&S}} , \label{opt semantic b}\\
    & p_{LB} \leq \rho_{k} \leq 1,  \forall k. \label{opt semantic c}
\end{align}
\end{subequations}

Optimisation problem \eqref{opt semantic} is convex and can be solved by using the bisection search method. 

The overall procedure for solving the optimisation problem \eqref{opt1} is outlined in Algorithm \ref{alg3}. In this algorithm, the stopping criterion is defined as $\left|\min_{k \in \mathcal{K}} \left( S_{k, e+1} \right) - \min_{k \in \mathcal{K}} \left( S_{k, e} \right) \right| \leq \varsigma$, where $S_{k,e}$ denotes the semantic secrecy rate at the $e$-th iteration for user $k$, and $\varsigma$ is a small predefined threshold that determines the convergence condition. When the difference between the minimum semantic secrecy rates in consecutive iterations is less than or equal to $\varsigma$, the algorithm stops, indicating that the optimisation has converged.

\begin{algorithm}
\caption{Alternating optimisation algorithm}\label{alg3}
\begin{algorithmic}[1]
\REPEAT
    \STATE With given $\mathbf{u}_{e}$ and $\rho_k$, solve optimisation problem \eqref{opt2}.
    \STATE Solve optimisation problem \eqref{opt FA} by using the objective function \eqref{log expansion obj}.
    \STATE Solve optimisation problem \eqref{opt semantic} by using the bisection search method.
    \STATE Update the epoch with $e = e+1$.
\UNTIL{Stopping criterion is satisfied}
\end{algorithmic}
\end{algorithm}

\section{Simulation Results}

In this section, we present numerical results to evaluate the effectiveness of the proposed design. The number of FAs and FPs are set to 5 and 7, respectively. The movable region of each FA is set to $0.0025 \text{m}^2$. The number of CUs and targets are set to 5 and 2, respectively. The power budget is set to 25 dBm, and the noise power is set to -30 dBm. 

\subsection{Semantic Communication}

\begin{figure}[!t]
    \centering
    \includegraphics[width=1\linewidth]{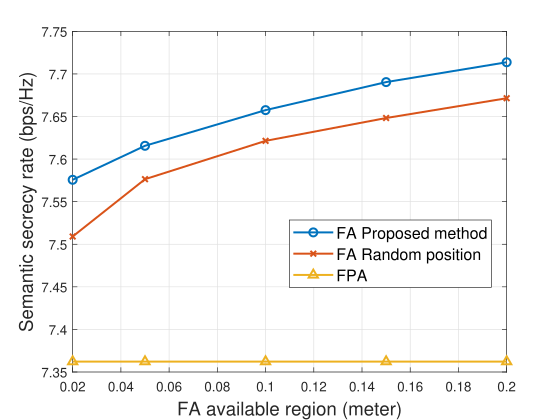}
    \caption{Achieved semantic secrecy rate against FAs movable region.}
    \label{fig gain vs ma}
\end{figure}

Fig. \ref{fig gain vs ma} illustrates the relationship between the semantic secrecy rate and the FA movable region for three different methods: the proposed method,  random position FAs, and FPA. The proposed method consistently achieves the highest semantic secrecy rate across all FA movable regions, demonstrating its superior performance. In contrast, the FA with random positions shows a moderate improvement as the FA movable region expands, although it remains inferior to the proposed method. Meanwhile, the FPA method maintains a constant and significantly lower semantic secrecy rate. The figure highlights that both the proposed method and FA with random positions benefit from an increased FA movable region, with the proposed method leveraging it more effectively.

\subsection{Sensing}

\begin{figure}[!t]
    \centering
    \includegraphics[width=1\linewidth]{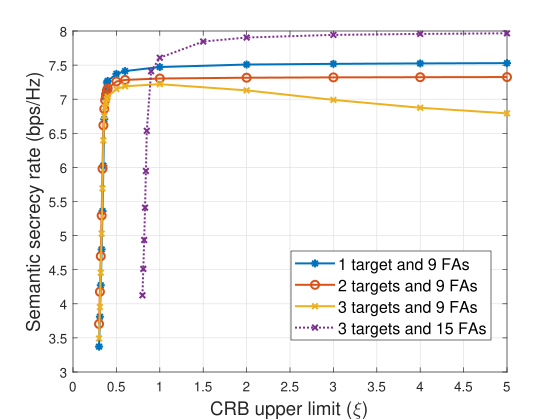}
    \caption{Worst-case semantic secrecy rate against CRB upper bound.}
    \label{crb sr}
\end{figure}

Fig. \ref{crb sr} illustrates the relationship between the worst-case semantic secrecy rate and the CRB upper limit, denoted as $\xi$, for varying numbers of targets and FAs. The CRB upper limit has been normalized with respect to the antenna size, as $\xi = \frac{\xi F}{\sigma_r^2 N_{\text{rx}} N_{\text{rz}}}$. 
The figure reveals a rapid increase in the semantic secrecy rate for small values of $\xi$, which subsequently reaches saturation or exhibits a slight decline as $\xi$ increases. This behaviour suggests that, beyond an optimal value of $\xi$, further increases in the CRB upper limit lead to no improvement in terms of communication performance, while also reducing sensing accuracy. A key observation is the impact of the number of targets. Configurations with fewer targets, such as "1 target and 9 FAs," achieve higher semantic secrecy rates compared to those with more targets, such as "3 targets and 9 FAs." This trend suggests that increasing the number of targets increases the likelihood of communication users being exposed to potential privacy breaches. The number of FAs also plays a critical role. Adding more antennas, as evidenced by the comparison between "3 targets and 9 FAs" and "3 targets and 15 FAs," consistently leads to an enhancement in the semantic secrecy rate. The addition of antennas likely provides better spatial diversity and enhanced sensing accuracy, helping to mitigate the challenges associated with a larger number of targets.

\section{Conclusion}


This paper introduces a novel joint design for FA-enabled ISCSC systems, where a BS can simultaneously communicate with multiple CUs and detect several extended targets. A joint optimization problem is formulated to maximize the worst-case semantic secrecy rate while ensuring critical constraints, such as sensing accuracy and power consumption, are met. To solve this non-convex optimization problem, it is decomposed into three sub-problems, which are addressed using an AO approach. The first sub-problem is tackled with a first-order Taylor expansion, ensuring convexity and simplifying the solution process. For the second sub-problem, a second-order Taylor expansion is applied to derive a surrogate function to approximate the objective function. The third sub-problem is resolved through a search-based method. Numerical simulations demonstrate the effectiveness of the proposed framework in enhancing semantic secrecy while maintaining the desired sensing performance within the power budget. These results underscore the potential of FA-enabled ISCSC systems to support the Metaverse.

\bibliographystyle{ieeetr}

\end{document}